\documentclass[journal=jpcafh,manuscript=article]{achemso}

\usepackage{chemformula} 
\usepackage[T1]{fontenc} 
\usepackage[utf8]{inputenc} %

\usepackage{siunitx}
\SectionNumbersOn

\author{Jamil Khalouf-Rivera}
\author{Miguel Carvajal}
\affiliation[University of Huelva]
{Depto. de Ciencias Integradas y Centro de Estudios Avanzados en Física, Matemáticas y Computación,  Universidad de Huelva, Huelva 21071, SPAIN}
\altaffiliation
{Instituto Carlos I de Física Teórica y Computacional, Universidad de Granada, Granada 18071, SPAIN}
\author{Lea F.\ Santos}
\affiliation[Yeshiva]
{Department of Physics, Yeshiva University, New York, New York 10016, USA}
\author{Francisco Pérez-Bernal}
\affiliation[University of Huelva]
{Depto. de Ciencias Integradas y Centro de Estudios Avanzados en Física, Matemáticas y Computación, Universidad de Huelva, Huelva 21071, SPAIN}
\altaffiliation
{Instituto Carlos I de Física Teórica y Computacional, Universidad de Granada, Granada 18071, SPAIN}
\email{curropb@uhu.es}
\phone{+34 959219789}
\fax{+34 959219777}

\title[]{Calculation of Transition State Energies in the HCN-HNC Isomerization  with an Algebraic Model}

\abbreviations{2DVM, ESQPT}

\begin{document}

\begin{abstract}
  Recent works have shown that the spectroscopic access to
  highly-excited states provides enough information to characterize
  transition states in isomerization reactions. Here, we show that
  information about the transition state of the bond breaking HCN-HNC
  isomerization reaction can also be achieved with the two-dimensional
  limit of the algebraic vibron model. We describe the system's
  bending vibration with the algebraic Hamiltonian and use its
  classical limit to characterize the transition state.  Using either
  the coherent state formalism or a recently proposed approach by
  Baraban \emph{et al.}  [\emph{Science} \textbf{2015}, \emph{350},
  1338-1342], we obtain an accurate description of the isomerization
  transition state. In addition, we show that the energy level
  dynamics and the transition state wave function structure indicate
  that the spectrum in the vicinity of the isomerization saddle point
  can be understood in terms of the formalism for excited state
  quantum phase transitions.
\end{abstract}


\section{Introduction}
Transition state theory is the keystone of chemical
reaction studies and chemical kinetics since its formulation in the
1930's\cite{Eyring1935,Wigner1938}. It allows for the derivation of
thermal reaction rates from the energy surface landscape, in
particular, from the minimal energy pathway connecting reactants and
products. However, the experimental study of transition states is
hindered by the saddle structure of the phase space region they
inhabit. In recent works, Baraban \emph{et al. } \cite{Baraban2015}
and Mellau \emph{et al. } \cite{Mellau2016} presented an interesting
approach that allows for the characterization of the transition state
in isomerization reactions using spectroscopic data in the frequency
domain as an input.  The approach in both works is based on a
particular spectroscopic pattern: the appearance of a dip in the
spacing of adjacent quantum levels for overtone series associated with
degrees of freedom that are connected with the reaction coordinate.

In molecular spectroscopy, the increase in the level density, which
happens together with the decrease in the energy difference between
neighboring energy levels, indicates that the system is reaching a
region subject to significant changes in its potential shape. An
example is the well-known energy level piling up that occurs once a
system gets close to dissociation. Birge and Sponer were already aware
of the importance of effective frequency\cite{Birge1926} and the
Birge-Sponer plots, of great relevance in the study of molecular
dissociation and in the estimation of dissociation energies, reflect
the decrease in the effective frequency value once the system
excitation energy approaches the dissociation energy. The deviation
from linearity in Birge-Sponer plots can be explained from the potential shape
and it may be parameterized to obtain a very precise estimation of the
system dissociation energy\cite{LeRoy1970, Leroy1970b}.

Nonrigid molecular systems also experience an increase in the energy
level density when the system explores the top of the barrier to
linearity. In this case, the adjacent quantum level splittings pass
through a minimum value, where anharmonicity switches from negative to
positive values. This feature is known as the Dixon dip since the
seminal work of Dixon, who showed that the vibrational bending degrees
of freedom of a quasilinear molecule can be modeled with a
cylindrically symmetric potential with a hump in the
origin\cite{Dixon1964}. He evinced the cusp in the effective frequency
at the energy of the barrier to linearity. This has been later
explained with the concept of quantum monodromy, which elucidates the
spectral features associated with the qualitative change in the system
phase space configuration that happens once the system energy reaches
the local potential maximum, at the top of the barrier to linearity
\cite{Child1998, Child1999, Child2008}.

The formalism presented in Ref.~\cite{Baraban2015} relates the dip in
the quantum levels spacing in isomerizing systems with the saddle (or
local maximum) structure of the potential at energies around the
isomerization barrier height.  The work proposed a simple
phenomenological formula to extract the isomerization barrier height
from spectroscopic data. This formula was applied to the
vibrational bending spectrum of two isomerizing systems that have been
subject to extensive theoretical and experimental analyses: the
HCN-HNC bond-breaking system and the \emph{cis}-\emph{trans}
configurations in the acetylene $S_1$ electronic state. In both cases,
it was shown how the proposed method helps to identify isomerization
pathways, allowing for the distinction between spectator vibrational
modes and those particular combinations of modes that favor the
isomerization reaction path. They were also able to extract the
transition state energies, $E_{TS}$, well within \(1\%\) of the value
of the isomerization barrier height obtained with sophisticated \emph{ab
  initio} calculations \cite{Harris2006,Mellau2010,Mellau2011,
  Baraban2016}.

Therefore, both isomerizing and nonrigid molecular species may be
described from a common perspective.  In the first case, the critical
point is associated with the transition state saddle point
\cite{Baraban2015, Mellau2016}, while in the second case, it is
connected with the top of the barrier to linearity
\cite{Winnewisser2006,Larese2011, Larese2013}. The questions addressed
here are whether an algebraic model like the two-dimensional limit of the vibron
model (2DVM) can be of help in the estimation of the
transition state properties from spectroscopic data and if the
decrease in the separation between energy levels can be considered 
as a new example of excited state quantum phase transition (ESQPT).

We show that the 2DVM, once furnished with
enough spectroscopic information, can also be used to characterize the
transition state with great accuracy. The 2DVM stems from the vibron
model introduced by Iachello in the 1980's, an algebraic model for
molecular structure that treats rovibrational excitations as bosonic
particles (vibrons) with a $U(4)$ dynamical algebra (or spectrum
generating algebra) \cite{Iachello:81, bookmol}. The 2DVM is 
tailored for the treatment of bending dynamics with a $U(3)$ dynamical
algebra \cite{Iachello1996}. Despite its apparent
simplicity, the 2DVM encompasses in a common framework the two
limiting cases of interest in the case of bending vibrations; rigidly-linear and rigidly-bent
configurations, as well as the feature-rich non-rigid case, with
particular spectroscopic signatures due to the existence of a barrier
to linearity \cite{Iachello:2003,
  PBernal2005}.
The classical (or mean-field)
limit of the 2DVM  can be obtained with the coherent
(or intrinsic) state formalism \cite{Gilmore1978,Gilmore1979}, which
provides an exact energy functional in the large system size limit \cite{PBernal2008}.

We perform calculations for the HCN-HNC system defining an algebraic
Hamiltonian and, in order to get close enough to the isomerization
barrier, we perform a fit to spectroscopically assigned \emph{ab
  initio} term values \cite{Harris2006}, as in
Ref.\cite{Baraban2015}. The assignment of these levels to the right
quantum labels was performed in Refs.~\cite{Mellau2010, Mellau2011}
and implied an exhaustive analysis of the full experimental
rovibrational spectrum for the [H,C,N] system. From the results of
this fit, we estimate the isomerization barrier energy in two ways. First, we obtain the energy functional associated with the
optimized algebraic Hamiltonian for both molecules making use of the
coherent state formalism. From the potential shape, we estimate the
value of the transition state energy. Next, we apply the
phenomenological formula put forward by Baraban \emph{et al.}
\cite{Baraban2015} to the term values predicted by the optimized
algebraic Hamiltonian, obtaining a second estimation of the saddle point energy. Moreover, we explore how the structure of the
wave function is affected once the system reaches energies around the
isomerization barrier and we relate this change to the occurrence of
an ESQPT in isomerizing systems.

The present work is organized as follows. Section \ref{th} gives a
brief introduction to the 2DVM algebraic formalism and on how to obtain
 its classical limit making use of the intrinsic (coherent) state
formalism. Section \ref{resd} presents and explains the results. Section \ref{con} contains our concluding remarks.

\section{Theory}
\label{th}
The modeling of $n$-dimensional many-body systems using a $U(n+1)$ spectrum
generating algebra  provides an effective description of a
large variety of systems \cite{bookalg}. The most successful examples
of this approach, undoubtedly, are the interacting boson model in
nuclear physics and the vibron model in molecular physics. The first
one is based on a $U(6)$ Lie algebra as its dynamical algebra
\cite{booknuc} and the second one relies on a $U(4)$ Lie algebra
\cite{bookmol}. In the present section, we briefly outline the
theoretical basis of the 2DVM.

The 2DVM was initially presented by
Iachello and Oss for the study of single and coupled benders
\cite{Iachello1996}. The model was found capable of reproducing the
characteristic spectroscopic features that plague the bending spectrum
of nonrigid molecular species \cite{Iachello:2003,
  PBernal2005}. Despite its apparent simplicity, the model includes
both a ground state and an excited state quantum phase transition.  By
conveniently parameterizing the 2DVM Hamiltonian, the system ground
state can be made to evolve from a rigidly-linear to a rigidly-bent
configuration through the variation of a control parameter. In this
process, the ground state undergoes a particularly abrupt change at a
critical value of the control parameter. This sudden change has been
interpreted as a quantum phase transition \cite{PBernal2008}, a
zero-temperature phase transition purely due to quantum fluctuations,
in the same fashion as in other many-body bosonic systems
\cite{Iachello2004,Cejnar2009,Cejnar2010}. ESQPTs, defined later \cite{Cejnar2006, Caprio2008},
generalize this concept to encompass excited states and are
characterized by a singularity (in the mean field limit) in the system
density of states at a critical energy value. This singularity defines
a separatrix between states having different character
\cite{Cejnar2006, Caprio2008}. Precursors of ESQPTs have been
identified in the vibrational bending spectra of several molecular
species and have been associated, through the intrinsic state
formalism, with the existence of a barrier to linearity in the energy
potential \cite{Larese2011, Larese2013}. The singularity in the
spectrum, marked by a pronounced decrease in level distance, happens
once the system energy approaches the top of the barrier.  The
particular spectroscopic features that appear at such energies were
explained introducing the concept of quantum monodromy
\cite{Child1998, Child1999}.  The development of new spectroscopy techniques has
made it possible to access experimentally excited vibrational states
at energies beyond the barrier to linearity
\cite{Winnewisser2005,Zobov2005}.  Quantum monodromy can be
interpreted as an ESQPT in the 2DVM \cite{Caprio2008, PBernal2008}, and
the spectral signatures found in the vibrational bending
spectra of some non-rigid molecular species have been considered the first
experimental confirmation of the occurrence of an ESQPT \cite{Larese2011,
  Larese2013}. Other experimental systems where ESQPT signatures have
been identified are superconducting microwave billiards
\cite{Dietz2013} and spinor Bose-Einstein condensates \cite{Zhao2014}.

\subsection{Algebraic approach to bending vibrations}

The algebraic approach to bending vibrations is based on a bosonic $U(3)$ Lie algebra, due to the inherently  2D nature of bending vibrations. The building bricks for this two-level boson model are  a scalar
boson, $\sigma^\dagger$, and two degenerate Cartesian bosons, $\{\tau^\dagger_x, \tau^\dagger_y\}$. The
nonzero  commutation relations between creation and annihilation operators
  are
\begin{equation}
\begin{array}{ccccc}
 \left[\sigma, \sigma^\dagger\right] = 1 &,& 
 \left[\tau_i, \tau^\dagger_j\right] = \delta_{i,j}   &;&i,j = x, y~.\\
\end{array}
\end{equation}
\noindent All other commutators are zero. It is convenient to
transform Cartesian into circular bosons \cite{PBernal2008}
\begin{equation}
\label{circ}
\tau^\dagger_{\pm} = \mp\frac{\tau^\dagger_x\pm i \tau^\dagger_y}{\sqrt{2}}~,~~
\tau_{\pm} = \mp\frac{\tau_x\mp i \tau_y}{\sqrt{2}}~.
\end{equation}

The nine $U(3)$ Lie algebra generators are the bilinear products of a
creation and an annihilation operator.  For a better physical insight,
they are expressed as \cite{Iachello1996, bookalg}:
\begin{equation}
  \begin{array}{lclcl}
    \hat n = \tau^\dagger_+\tau_++\tau^\dagger_-\tau_- &  , &
    \hat n_s = \sigma^\dagger\sigma &,&   \hat \ell = \tau^\dagger_+\tau_+-\tau^\dagger_-\tau_-~,  \\
    \hat D_\pm = \sqrt{2}(\pm\tau^\dagger_\pm\sigma\mp\sigma^\dagger\tau_\mp)&  , & 
    \hat Q_\pm = \sqrt{2}\,\tau^\dagger_\pm\tau_\mp &,&
    \hat R_\pm =\sqrt{2} (\tau^\dagger_\pm\sigma+\sigma^\dagger\tau_\mp)~.
  \end{array}
  \label{gen}
\end{equation}

The next step in the algebraic procedure is to consider the possible dynamical symmetries, subalgebra chains starting in the dynamical algebra and ending in the
system's symmetry algebra. In
the present case,  the system is
limited to a plane and 2D angular momentum (vibrational angular
momentum in molecular bending vibrations) is conserved; thus the symmetry algebra is the $SO(2)$ Lie. The generator of  $SO(2)$ is the vibrational angular momentum, \(\hat \ell\), as one can easily understand
once it is expressed in terms of the Cartesian boson operators. As this is
an angular momentum projection in the direction perpendicular to the
system's plane, it can take both positive and negative (or zero)
values. There are two possible dynamical symmetries that start in
$U(3)$ and end in $SO(2)$
\begin{subequations}
\label{chains}
\begin{eqnarray}
&U(3)\supset U(2)\supset SO(2) &\mbox{Chain I}~,\label{chaini}\\
&U(3)\supset SO(3)\supset SO(2) &\mbox{Chain II}~.\label{chainii}
\end{eqnarray}
\end{subequations}

Each dynamical symmetry conveys a basis and an analytical energy
formula that can be associated with a physical limiting case. The $U(2)$ dynamical symmetry, also called the cylindrical oscillator symmetry, corresponds to a rigidly-linear molecule; while the $SO(3)$ dynamical symmetry is associated with a rigidly-bent configuration.  A detailed discussion of both dynamical
symmetries, their geometric implications, and the relation between them can be found in \cite{PBernal2008}. All calculations in the present work have been performed using the chain I basis, the cylindrical oscillator basis, whose states are denoted as  $|[N];n^\ell\rangle$. The quantum number $N$ labels the totally symmetric representation of $U(3)$ and it is
related to the total number of bound states of the system. Being a constant, hereafter we simplify the basis states notation to $|n^\ell\rangle$. The quantum label $n$ indicates the
vibrational number of quanta, and $\ell$ is the vibrational angular momentum. The
 branching rules are 
 \begin{eqnarray}
 n & = & N, N-1, N-2, \ldots, 0 \nonumber\\
 \ell & = & \pm n, \pm (n-2), \ldots, \pm 1 \mbox{ or }0~,~~ (n = \mbox{odd or
   even}) ~.
 \end{eqnarray}

The definition of a simple Hamiltonian that contains the main
physical ingredients of the model and allows for the study of the shape phase transition between the different dynamical symmetries implies the consideration of Casimir
or invariant operators of the subalgebra chains under study \cite{Gilmore1978,Gilmore1979}. A simple model Hamiltonian includes the first order Casimir operator of $U(2)$, $\hat n$, and the second order Casimir
operator of $SO(3)$, $\hat W^2 =(\hat D_+\hat D_-+\hat D_-\hat D_+)/2+\hat \ell^2$.

To reproduce the bending spectrum of HCN and HNC, we use the algebraic Hamiltonian

\begin{equation}
    \label{HAM}
    \hat H = P_{11} \hat n + P_{21}\hat{n}^2 + P_{22} \hat{\ell}^2 + P_{23}\hat{W}^2 + P_{45}\left[\hat{W}^2\hat{n}^2+\hat{n}^2\hat{W}^2\right]~,
\end{equation}
\noindent extending the most general one- and two-body Hamiltonian, employed in
\cite{Larese2013}, with a four-body operator.  The parameter $P_{ij}$ comes with the $j$-th $i$-body operator in the Hamiltonian. The matrix elements of the one- and two-body operators in chain I basis are
\begin{align}
  \langle n^\ell|\hat n|n^\ell\rangle &= n~,~~ \langle n^\ell|\hat n^2|n^\ell\rangle = n^2~,~~ \langle n^\ell|\hat{\ell}^2|n^\ell\rangle = \ell^2~,\\
  \langle n_2^\ell|{\hat W}^2|n_1^\ell\rangle &=- \sqrt{(N-n_1+2)(N-n_1+1)(n_1+\ell)(n_1-\ell)}\,\delta_{n_2,n_1-2}\nonumber\\
  &+\left[(N-n_1)(n_1+2)+(N-n_1+1)n_1 + \ell^2\right] \delta_{n_2,n_1}\\
  &-\sqrt{(N-n_1)(N-n_1-1)(n_1+\ell+2)(n_1-\ell+2)}\,\delta_{n_2,n_1+2}~.\nonumber
\end{align}

The four-body operator \(\left[\hat{W}^2\hat{n}^2+\hat{n}^2\hat{W}^2\right]\) is  used here exclusively to improve the HCN data fit, and its matrix elements are

\begin{align}
\langle n_2^\ell|\hat n^2 \hat W^2 + \hat W^2 \hat n^2|n_1^\ell\rangle = &-[n_1^2 + (n_1-2)^2]\sqrt{(N-n_1+2)(N-n_1+1)(n_1+\ell)(n_1-\ell)}\,\delta_{n_2,n_1-2}\nonumber\\
&+ 2n_1^2\left[(N-n_1)(n_1+2)+(N-n_1+1)n_1 + \ell^2\right] \delta_{n_2,n_1} \label{W2n2matel_u2}\\
&- [n_1^2 + (n_1+2)^2]\sqrt{(N-n_1)(N-n_1-1)(n_1+\ell+2)(n_1-\ell+2)}\,\delta_{n_2,n_1+2}~~.\nonumber
\end{align}

Operators $ \hat n$, $\hat{n}^2$, and $\hat{\ell}^2$ are diagonal in the $U(2)$ basis and can be identified with a harmonic term, an anharmonic correction, and the vibrational angular momentum, respectively. By contrast, the operator \(\hat{W}^2\) is diagonal in the chain II basis and it is associated with an anharmonic displaced oscillator. The four-body operator combines Casimir operators from both subalgebra chains and it is not diagonal in any of them.  Using the procedure sketched in the next section, we obtain a set of optimized spectroscopic parameters $P_{ij}$ for each molecule.

\subsection{The classical limit of the two-dimensional vibron model}

A system energy functional can be retrieved from the algebraic
Hamiltonian (\ref{HAM}) by the method of coherent (intrinsic)
states originally introduced in the study of nuclei
\cite{Gilmore1978,Gilmore1979,Ginocchio1980,Dieperink1980}, and later
adapted to molecular systems \cite{Vanroosmalen1982}. There are other methods to establish a link between the phase space and the algebraic approaches \cite{SCastellanos2012}.

The intrinsic state method defines a coherent state, where the variational
parameters $r$ and $\theta$ are, in general, complex and
represent coordinates and momenta \cite{Vanroosmalen1982}. We consider
the spatial dependence only and, therefore, we set the momenta equal
to zero \cite{bookmol}. We now proceed to outline the more relevant
results needed to obtain the classical limit of the 2DVM. For a
detailed description of this procedure in the 2DVM case see
Refs.~\cite{PBernal2008,Larese2011, Larese2013}.

The first step is the  coherent state definition
\begin{equation}
|{\mbox{c.s.}} \rangle \equiv  |[N];r,\theta\rangle =
  \frac{1}{\sqrt{N!}}
  \left(b_{c}^\dagger\right)^{N}|0\rangle, 
\label{cohs}
\end{equation}
\noindent where $r$ and $\theta$ are the polar coordinates associated
with Cartesian coordinates $x$ and $y$. The operator $b_{c}^\dagger$ is the boson condensate creation operator, 
\(b_{c}^\dagger = \frac{1}{\sqrt{1+r^2}}\left[\sigma^\dagger +  
    \left(x\tau_{x}^\dagger + y\tau_{y}^\dagger\right)\right]\).

The expectation value  of the Hamiltonian (\ref{HAM}) in the
coherent state gives as a result  the ground state energy functional, $E(r)$, akin to the system potential function
\begin{align}
  {E}(r) =& \frac{\langle {\mbox{c.s.}}| \hat H  |{\mbox{c.s.}} \rangle}{N} \nonumber\\
  =& P_{11}\frac{r^2}{1+r^2}+  P_{21}\left[\frac{r^2}{1+r^2}+(N-1)\frac{r^4}{(1+r^2)^2}\right] + P_{22}\frac{r^2}{1+r^2}+P_{23}\left[2 +(N-1)\frac{4r^2}{(1+r^2)^2}\right]  \nonumber\\ 
          & + P_{45}\left[ 4\frac{r^2}{1+r^2}+(N-1)\frac{12r^4+16r^2}{(1+r^2)^2}\right.\label{efun}\\
  &\left.+ (N-1)(N-2)\frac{4r^6+28r^4}{(1+r^2)^3} + (N-1)(N-2)(N-3)\frac{8r^6}{(1+r^2)^4}\right]\nonumber
\end{align}

The equilibrium configuration of the molecule is obtained by
minimizing $E(r)$ with respect to the variable $r$.

\section{Results and Discussion}
\label{resd}

To cast light upon the estimation of the transition state properties and the possible link between  isomerization and ESQPTs,  we analyze the available
data for the HCN-HNC system. The available experimental data for the
bending vibrational spectrum of HNC and HCN were already successfully modeled with
a four-parameter 2DVM spectroscopic Hamiltonian, which is the most
general Hamiltonian including one- and two-body interactions \cite{Larese2013}. Unfortunately, experimental data are not available
above \SI{10000}{cm^{-1}} and the dissociation barrier is expected to
lie around \SI{17000}{cm^{-1}} above the HCN minimum, which is located
approximately \SI{5200}{cm^{-1}} below the HNC minimum. To
overcome this obstacle, we adopt the same approach as in Baraban
\emph{et al.}  \cite{Baraban2015}: we use for our calculations a  set of \emph{ab initio} term values \cite{Harris2006} 
spectroscopically assigned  after an exhaustive analysis of the full experimental rovibrational spectrum for the [H,C,N] system \cite{Mellau2010, Mellau2011}. In the present work, selecting pure bending levels, we consider 142
energies with vibrational angular momenta up to $\ell=12$ in the
case of HCN, compared to 30 available experimental terms; and 41
energy levels up to \(\ell = 9 \) in the case of HNC, compared
to only 19 experimental levels. The optimization of the spectroscopic parameters was carried out  through an iterative non-linear
least-square fitting procedure that uses the Fortran version of
\texttt{Minuit} \cite{minuit} and the optimal values computed are included in Tab.~\ref{Pij} (see supplementary text for details of the
fits).

\begin{figure}[H]
  \centering \includegraphics[trim=0 100 0
  100,width=0.7\textwidth,angle=-90]{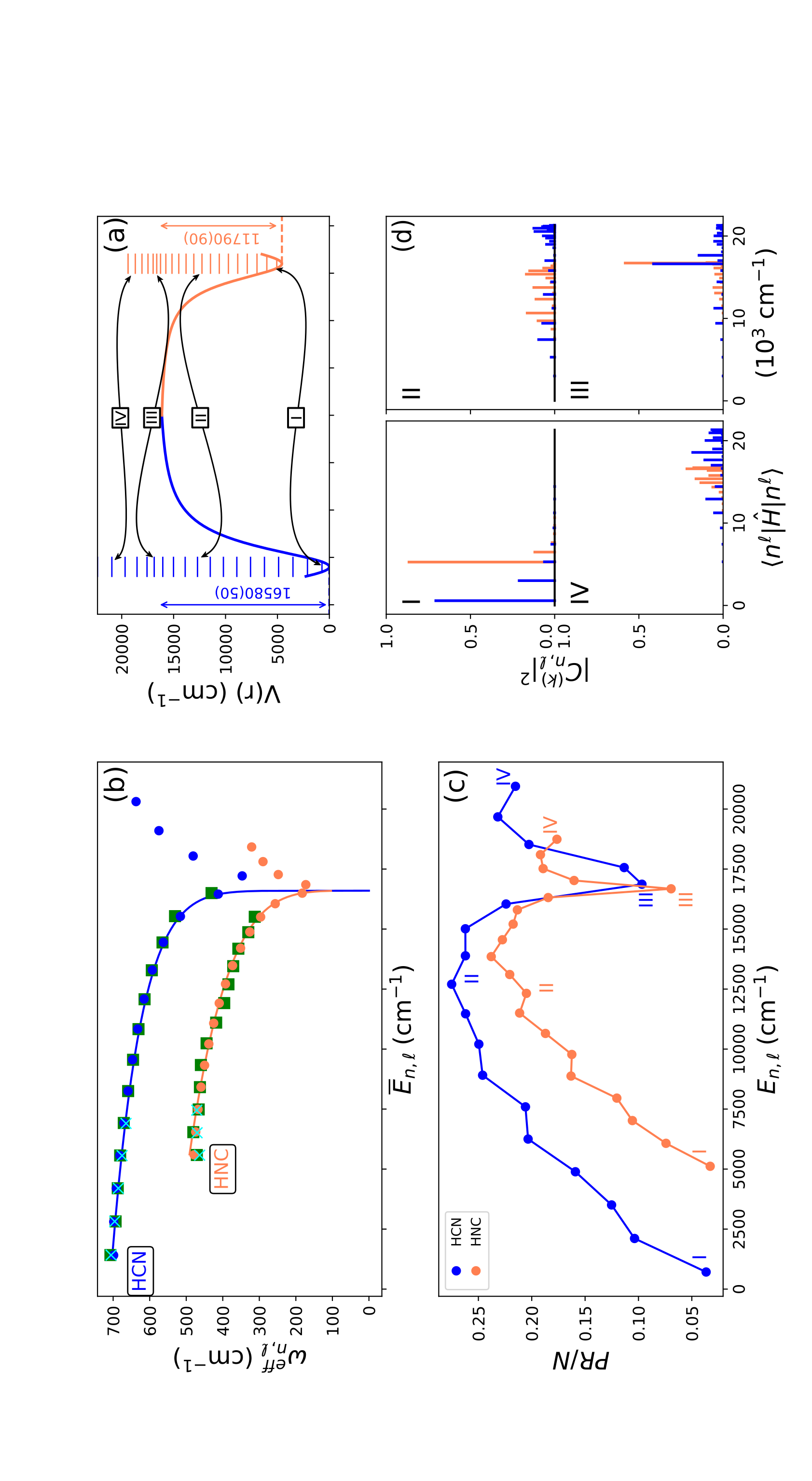}
    \caption{(Color online) Panel (a): sketch of the potentials
      obtained for the two molecular species using the intrinsic state
      formalism and the isomerization barrier values, locating the
      states I, II, III, and IV chosen to illustrate our results (see
      text).  Panel (b): effective frequency \(w^{eff}\) as a function
      of the midpoint excitation energy \(\overline E\) for HCN and
      HNC. Crosses indicate the available experimental data, while
      green squares are spectroscopically assigned \emph{ab initio}
      results (Ref.~\cite{Harris2006}, see text). Blue (orange)
      circles are the algebraic model results and the blue (orange)
      line marks the result of fitting Eq.~(\ref{Baraban}) to the
      algebraic fit results for HCN (HNC). Panel (c): Normalized
      participation ratio (see Eq.~\ref{PRdef}) of \(\ell = 0\) HCN
      (blue dots) and HNC (orange dots) eigenstates resulting from the
      fit of the algebraic Hamiltonian (\ref{HAM}) (See supplementary
      material for details) making use of a truncated cylindrical
      oscillator basis. Panels (d): Squared components in the
      cylindrical oscillator basis for four algebraic model
      eigenstates of HCN (blue bars) and HNC (orange bars) as a
      function of the expectation value of the Hamiltonian in the
      \(|n^\ell\rangle\) basis state.  }
    \label{elle0}
\end{figure}

Once we optimize the spectroscopic parameter values in the algebraic Hamiltonian (\ref{HAM}), we compute the vibrational bending energy functional for
both HCN and HNC using the Eq.~(\ref{efun}) derived from the intrinsic state
formalism.  The obtained functionals, depicted in Fig.~\ref{elle0}a, allow for the estimation of the  isomerization barrier height, \(E_{TS}\),
 which corresponds to the distance between the functional minimum and
its asymptotic value. We provide in the column labeled 2DVM-I of Tab.~\ref{Ets_com} the obtained \(E_{TS}\) values  for
HCN and HNC.

Baraban \emph{et al.} calculate the transition state energy with a different approach. They quantify the observed energy dip
making use of the \emph{effective frequency}, a quantity defined for
quantum systems as \(w^{eff}(n) = \frac{\partial E(n)}{\partial n} = \frac{\Delta E}{\Delta n}\), i.e. the
discrete derivative of the system energy with respect to the principal
quantum number \(n\) \cite{Baraban2015,Mellau2016}. They suggest a simple formula\cite{Baraban2015} to parameterize the dependence
of the effective frequency on the midpoint vibrational energy \(\overline E\),

\begin{equation}
 w^{eff}(\overline E) =  \omega_0 \left(1-\frac{\overline E}{E_{TS}}\right)^{\frac{1}{m}}~,
\label{Baraban}
\end{equation}
with three adjustable parameters: \(\omega_0\), \(m\), and
\(E_{TS}\). The parameter \(\omega_0\) is the effective frequency for
the potential ground state and \(m\) depends on the potential
shape\cite{Baraban2015}. The most relevant parameter is \(E_{TS}\), the
transition state energy, which provides an estimate of the energy
barrier between different reactants.

We  derive the transition state energy employing an alternative procedure. We
use the effective frequencies computed from the term values predicted using the 2DVM, for
both molecular species. The results for  \(\ell = 0 \) vibrational angular
momentum are shown in Fig.\ \ref{elle0}b, where
\(\omega^{eff}\) is plotted as a function of \(\overline E\) using
blue (orange) circles for the algebraic model results for HCN
(HNC). The effective frequency for spectroscopically assigned \emph{ab
  initio} data is depicted using green squares and the available
experimental data are also included as (cyan) crosses. It is clear
from this figure that the 2DVM results undergo the expected dip in the
effective frequency and that they provide a very good estimate of
the height of the isomerization barrier in the HCN-HNC molecular
system. The HNC data are displaced so that the top of the barrier is
common for both molecular species, which allows for the estimation of
the separation between the HCN and HNC energy minima. We use the
effective frequency of the algebraic term values to fit the parameters
in the function (\ref{Baraban}) with the help of the Python
\texttt{LMFIT} package \cite{lmfit}, and obtain the estimated barrier
values in the column of Tab.~\ref{Ets_com} labeled as 2DVM-II.

The agreement between the values of the transition state energy
obtained with the two methods above and the values obtained with other approaches is very good, as seen in
Tab.~\ref{Ets_com}. The differences of the \(E_{TS}\) value with respect to sophisticated \emph{ab initio}
calculations are of 1-2\% in the HCN case, and they increase to a maximum
of 4\% in the HNC case. The explanation for this difference lies in the
fact that the HNC bending potential has an unusual shape from the
interaction with a nearby excited diabatic electronic state
\cite{Lauvergnat2001}. In order to overcome this obstacle, Baraban
\emph{et al.}\cite{Baraban2015} included a Gaussian term to their
phenomenological formula (\ref{Baraban}). The lack of this extension
in the present work explains the different agreement with the results
obtained using other approaches for HCN and HNC.

\begin{table}[H]
    \centering
    \begin{tabular}{|r||c|c|c|c|c|c|c|}
    \hline    
      Molecule & \(P_{11}\) & \(P_{21}\) & \(P_{22}\) & \(P_{23}\) & \(P_{45}\times10^4\) & \(N\) & \(rms\) \\ 
      \hline \hline
      HCN &  2308.3(6) & -39.947(14) & 21.810(6) & -10.635(3) & -1.311(3) & 50 & 19.37\\
      \cline{2-8}
      HNC & 1024.9(1.4) & -18.59(4) & 13.362(23) & -5.085(11) & - & 40 & 14.91 \\
    \hline 
    \end{tabular}
    \caption{\label{Pij}Optimized spectroscopic parameters $P_{ij}$ (\si{cm^{-1}}  units), root mean square deviation \(rms\) (\si{cm^{-1}}), and vibron number \(N\) obtained from the fit to the HCN and HNC \emph{ab initio} data set \cite{Mourik2001,Harris2006,Mellau2011} for  Hamiltonian (\ref{HAM}). For a detailed description of the fitting procedure see the supplementary material.}
\end{table}

\begin{table}[H]
    \centering
    \begin{tabular}{|c||c|c|c|c|c|}
        \hline
        \multicolumn{6}{|c|}{\(E_{\text{TS}}\) comparison ($\text{cm}^{-1}$)}\\
        \hline
         Molecule        & 2DVM-I  & 2DVM-II     & Baraban \emph{et al.}~\cite{Baraban2015} & Mourik \emph{et al.}~\cite{Mourik2001} & Makhnev \emph{et al.}~\cite{Makhnev2018} \\
         \hline \hline
        HCN($\ell=0$)    & 16580(50) & 16599(15)   & 16695(17)                                & 16798                                  & 16809.4 \\
        HNC($\ell=0$)    & 11790(90)  & 11977(15)   & 11533(124)                               & 11517                                  & 11496.6 \\

        \hline
    \end{tabular}
    \caption{\label{Ets_com}Isomerization barrier height in \si{cm^{-1}}  units for HCN and HNC computed from the bending energy functional (2DVM-I) and from the 2DVM optimized term values  using Eq.\ (\ref{Baraban}) (2DVM-II) compared with results obtained using other approaches (columns third to fifth).}
\end{table}

An advantage of the algebraic model, compared to a pure Dunham
expansion, is that it provides not only the spectrum, but also the eigenstates of 
 Hamiltonian (\ref{HAM}) upon diagonalization. The height of the
isomerization barrier has strong consequences for the structure of the eigenstates and the dynamics of the HCN-HNC
molecular system \cite{Bacic1987, Light1987, Mellau2011b}. Indeed, it has been recently shown that the system's eigenstate with
the closest energy to the saddle point that characterizes the
transition state has an enhanced localization in the bending
coordinates \cite{Mellau2016}. A similar phenomenon has been discussed
in the case of ESQPTs in different realizations of the vibron
model, where eigenstates with eigenvalues close to the critical energy
of the ESQPT have been shown to be strongly localized in the basis associated
with the linear configuration, also called the cylindrical oscillator
basis \cite{Santos2015, Santos2016, PBernal2017}.

The level of localization of states written in a certain basis can be
quantified with quantities such as the information (or Shannon) entropy or the participation ratio (PR) \cite{Zelevinsky1996, Gubin2012}. A large PR value implies that the state receives significant contributions from many basis states, and a small PR value denotes a strong localization of the state in the basis. This is similar to other quantities, as Heller's F parameter, used to probe phase space flow in molecular systems \cite{Heller1983, Heller1987}.
Hamiltonian (\ref{HAM}) is block-diagonal in the vibrational angular
momentum \(\ell\). Its eigenstates can be written as a linear
combination of the cylindrical oscillator basis states
\(\{|[N]n^\ell\rangle\}\):
\(|\psi^{(\ell)}_{k}\rangle = \sum_n C^{\ell}_{k,n}|n^\ell\rangle\)
and the participation ratio is defined as
\begin{equation}
  \text{PR}\left(|\psi^{(\ell)}_{k}\rangle\right) = \left[\sum_{n} \left|C^{\ell}_{k,n}\right|^4\right]^{-1}~.
  \label{PRdef}
\end{equation}
The minimum PR value is \(1\), when the system localization is maximal
and the eigenstate can be identified with a basis state. The maximum
PR value is the dimension of an \(\ell\)-vibrational angular momentum
basis block, which happens when all components are nonzero and have
the same weight.

 In the nonrigid
case, the eigenstate at the ESQPT critical energy is strongly localized  in the Chain I basis state with \(n = 0\). This fact has been shown to affect the 
system dynamics. If the system is initially prepared in this initial basis state, the evolution is substantially slower than for other initial states with similar energy \cite{Santos2015, Santos2016, PBernal2017, Kloc2018}. Similarly, the eigenstate at the isomerization barrier is also localized, but with a caveat, as we explain next. In the HCN-HNC case,
we plot the PR for the \(\ell = 0\) eigenstates normalized by the
vibron number \(N\) in Fig.~\ref{elle0}c with blue (orange) dots for
HCN (HNC). In both cases, there is a remarkable decrease in the PR
value for eigenstates close to the isomerization energy value. To further clarify the variation in the eigenstates structure,
we include in Fig.~\ref{elle0}a the two energy functional curves and
the energies of four eigenstates chosen to illustrate the different
structures of the wave functions at different excitation energies. These states are also indicated in Fig.~\ref{elle0}c. The
four selected eigenstates of HCN and HNC, labeled as \(I\), \(II\),
\(III\), and \(IV\), have energies at different locations in the potential:
at the ground state energy, at a mid height, at the isomerization barrier,
and above the barrier.  The bar diagrams in the four panels of
Fig.~\ref{elle0}d are drawn from the squared value of the
\(C^{\ell}_{k,n}\) coefficients for those four eigenstates  as a function of the energy of the corresponding basis
vector, \(\langle N n^\ell|\hat H|N n^\ell\rangle\); bars are blue
(orange) for HCN (HNC) eigenstates. The change in the structure as we move from eigenstate \(I\) to \(IV\) is evident. The low energy eigenstate \(I\) is localized, because
the cylindrical oscillator basis is the most appropriate basis for the
description of rigidly-linear configurations. Eigenstates  \(II\) and
\(IV\) are characterized by a strong mixing in the same basis. Eigenstate
\(III\) is of special relevance, since it is the eigenstate with the
closest energy to the isomerization barrier. It is characterized by a
strong localization in the cylindrical oscillator basis. Contrary to
what happens in the ESQPT for nonrigid molecules, where for
\(\ell = 0 \)  the basis
element \(|N 0^0\rangle\) has the largest component \cite{Santos2015, Santos2016, PBernal2017},
in the isomerization case, the minimum in the PR is associated to a
large component in a basis state with a high \(n\) value, e.g.
\(|N N^0\rangle\) for an even \(N\) value. This is likely caused by anharmonicity effects in the Hamiltonian, as already hinted
in Ref.~\cite{PBernal2010}. Indeed, we verified that a negative
quadratic contribution in the vibrational quantum number operator has effects in the
symmetric phase for linear and quasilinear states before the
control parameter reaches the critical value.

\section{Conclusions}
\label{con}
In short, we have  shown that the 2DVM, despite the simplicity of the Hamiltonian (\ref{HAM}), describes extremely well the localization and the effective frequency dip  of the transition state for
isomerizing systems, once it is fed with enough spectroscopic data or
with accurate enough \emph{ab initio} calculations. The value of the
transition state energy can be estimated from the intrinsic state
energy functional or from the dip in the energy gap. In both cases the
differences with the \(E_{TS}\) values obtained with sophisticated \emph{ab initio}
calculations are very small, in the range 1-4\%.  As a consequence of the link between the isomerization barrier and ESQPT's, our characterization of the transition state is not restricted to energy values and their differences only, but it includes also the structure of the algebraic wave function. This offers a
promising line of research for applications of the ESQPT formalism to
isomerization reactions.

Our approach also provides a physically sound way for
obtaining a minimum bound for the vibron number value \(N\), which
needs to be large enough to accommodate the minimum in the
participation ratio. Heretofore, in the case of bending vibrations, the value of the
vibron number \(N\) used to be fixed based only on the best fit to
experiment \cite{Larese2011}.

Another problem that can be tackled with the present formalism is the isomerization between the \emph{cis} and \emph{trans} geometric configuration of acetylene in the acetylene $S_1$ electronic state. The modeling  in the algebraic framework of the bending degrees of freedom in a tetratomic molecule, which implies two coupled benders, requires a $U(3)\otimes U(3)$ dynamical algebra \cite{Iachello1996,Champion1999,Iachello:2009,Larese2014}. An appropriate starting point would be the fit to the experimental bending vibrational term levels for each one of these two acetylene geometric configurations obtained making use of an algebraic Hamiltonian based on coupled dynamical algebra $U(3)\otimes U(3)$ \cite{Larese2014}. The study of this case is of interest because this would be the first example of the identification of an ESQPT  in experimental data for a system with more than one effective degree of freedom. In such systems, the detection of the ESQPT precursors is expected to be  more cumbersome than in the single degree of freedom cases \cite{Stransky2014,Stransky2015}.

Finally, it is interesting to note that the 2DVM eigenstates with
positive slope in the right end of Fig.~\ref{elle0}b have energies beyond the isomerization energy
barrier. Thus, they cannot be unambiguously associated to one of the two
molecular configurations and they correspond to the so-called bond-breaking states,
which are often expressed with the H\(_{0.5}\)CNH\(_{0.5}\)
formula\cite{Mellau2011b}. These eigenfunctions necessarily entangle both
molecular configurations, something that in the 2DVM case was already noticed in
Ref.~\cite{Larese2013} (see Fig.~6 in this reference). A full
description of the isomerizing HCN-HNC system at these energies requires the
consideration of both molecules in a single system. This is a direction we plan to explore in the near future.

\begin{acknowledgement}


  The authors thank José Miguel
  Arias,  José Enrique García-Ramos, Franco Iachello, Georg Mellau, and Pedro Pérez Fernández for useful discussions and comments. JK thanks support from the Youth
  Employment Initiative and the Youth Guarantee program supported by the
  European Social Fund.  LFS is supported by the NFS Grant No. DMR-1603418. This study has been partially financed by the
  Consejería de Conocimiento, Investigación y Universidad, Junta de Andalucía
  and European Regional Development Fund (ERDF), ref. SOMM17/6105/UGR and by
  the Centro de Estudios Avanzados de Física, Matemáticas y Computación
  (CEAFMC) of the Universidad de Huelva. Computer resources supporting
  this work were provided by the CEAFMC and Universidad de Huelva High
  Performance Computer (HPC@UHU) located in the Campus Universitario
  el Carmen and funded by FEDER/MINECO project UNHU-15CE-2848.
\end{acknowledgement}

\begin{suppinfo}



The following files are available free of charge as supporting information.
\begin{itemize}
\item \texttt{Sup\_Mat.pdf}: A brief explanation of the procedure
  followed to obtain the classical limit of the 2DVM and of the
  effective frequency fitting procedure. Results of the model for non-zero
  vibrational angular momentum.
\item \texttt{SM\_energies.txt}: Results obtained from the 2DVM fit
  to spectroscopically assigned \emph{ab initio} energies including the comparison with experimental
  data when available.
\end{itemize}

\end{suppinfo}

\bibliography{achemso-jpclett-revised}

\begin{tocentry}
\includegraphics[angle=0,width=0.9\linewidth]{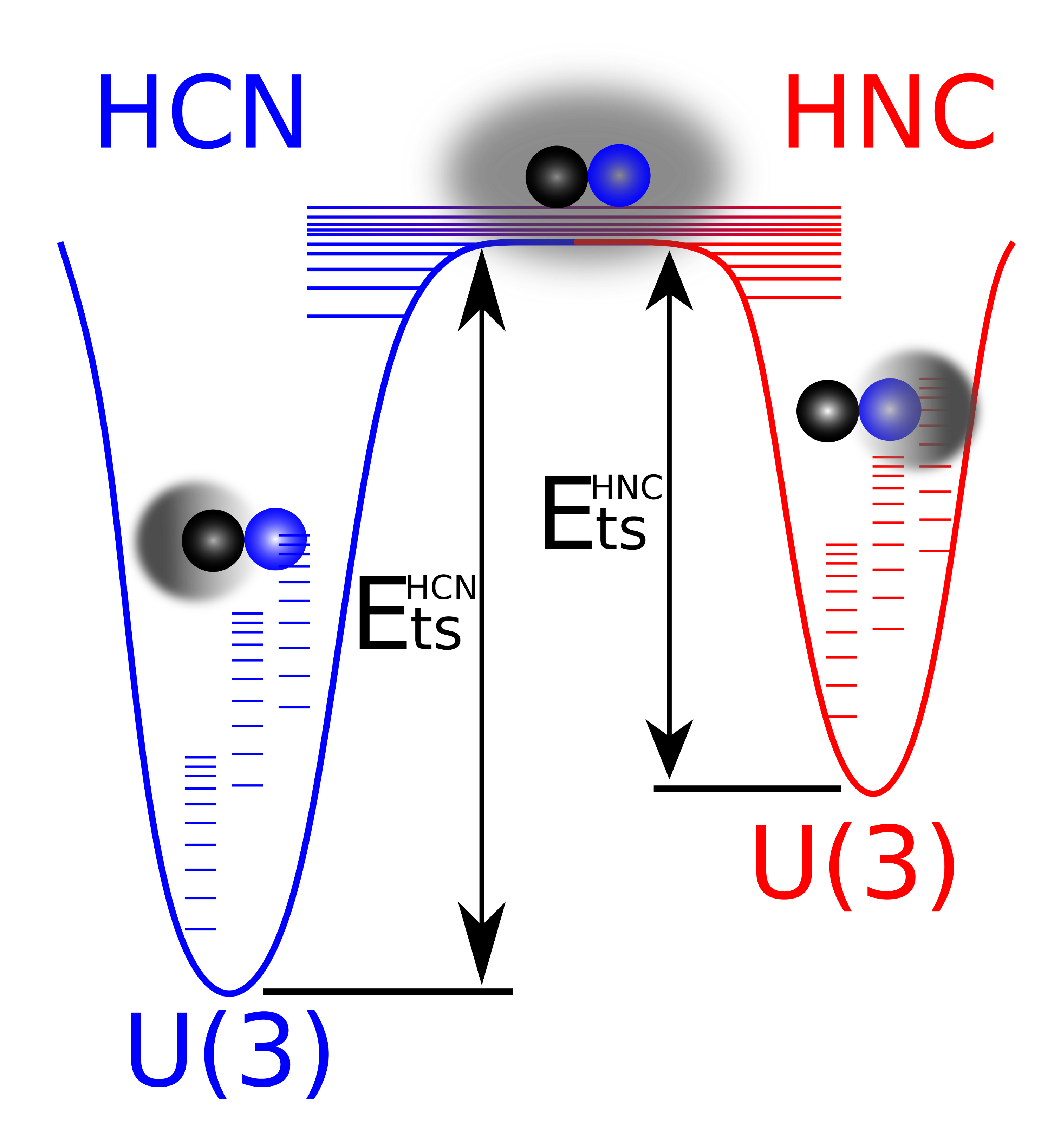}
\end{tocentry}

\end{document}